# Phonon tunnels across a sonic horizon

H. Z. Fang[1] and K. H. Zhou[2]

*Department of Physics, China University of Petroleum, Changjiang Road 66, 266555, Qingdao, China*

Abstract: We consider phonon tunneling in sonic black hole by WKB approximation method without the backreaction, in which the relativistic momentum-energy equation is used. We also study the WKB loop (closed path) integral [B. D. Chowdhury, Pramana 70:3-26, 2008] in sonic black hole case and discuss the equivalence of the results of some cases from the two different tunneling probability expression in Painlevé-like coordinates.

## 1. Introduction

Particle radiation from the black hole horizon [1, 2] is widely studied in different ways, but there is no experimental confirmation yet (maybe the relative experiment in analog system will carry out in the near future [3]) and some problems (i.e. the paradox of information loss and trans-Planckian problem) stay open. Along the way to derive the Hawking radiation, the consistent results are obtained but different features are emphasized under different assumptions. The Hartle-Hawking approach used analytic continuation of the propagator across the event horizon of an eternal black hole [4]. Damour and Ruffini emphasized that the outgoing modes suffer a barrier when crossing the horizon [5]. Robinson and Wilczek claimed that there exists relationship between Hawking radiation and gravitational anomalies [6] and some physicists believe that Hawking effect can be interpreted in string theory [7].

One interesting work [8] brings the semi-classical WKB method into calculating the Hawking radiation which gives a more intuitive and reasonable picture of the particle tunneling through the horizon. When the energy conservation and self-gravity are considered, the result from WKB method supports that the information are conserved in the process of black hole evaporation, which is consistent in several black hole spacetimes [9-14]. It should be pointed out that the WKB tunneling amplitude is not invariant due to the classical particle action is various in canonical transform, for which, instead, one should use the modified formula: WKB loop integral [15]. But the WKB loop integral gives twice the Hawking temperature. Including the time jump at the horizon with an imaginary part for that the properties of space-time coordinate exchanges when crossing the horizon, the expected result is recovered [16-18]. That is to say both the temporal contribution and spatial contribution should be counted in the WKB loop integral.

Unruh took the first step in analogy of/for gravity by the derivation to the special sonic effective metric and phonon emission from the horizon [19], which brings many opportunities and new research topics for experimentally detecting Hawking radiation, shedding light on quantum gravity and getting new viewpoints of some known physics. In the analogue gravity field, the Hawking radiation from various "horizons" is also studied [20], which even contain different dispersion relationships [21, 22]. In sonic black hole study, the semi-classical WKB method is involved in some way [23, 24]. Here we will apply it to calculate the sonic Hawking radiation and attempt to find some insights about the WKB method in sonic black hole evaporation process.

---

[1] Email address: qdfang@163.com
[2] Email address: kaihu558895@sina.com

In this paper, we apply both the classical WKB method and the modified WKB method to analyze the sonic Hawking radiation in spherically symmetric sonic black hole and two-dimensional rotational sonic black hole under the classical background, then based on which we prove that both the classical momentum integral and the WKB loop integral in 1-D and 2-D stationary spacetime cases give the same tunneling probability across the horizon in Painlevé-like coordinate system.

## 2. Phonon tunneling in classical WKB approximation

In acoustic analogy system, for barotropic, viscid-free, and irrotational fluid flow, the sound wave propagating in the background is similar to the light travelling in curved spacetime, which can be controlled by an effective metric. When the flow velocity becomes transonic, the sonic horizon will appear, which is determined by the condition $v_\perp = c_s$ (where $v_\perp$ is the normal component of the flow velocity $v$ in respect to the horizon and $c_s$ is the local sound speed) [25]. As used in gravity particle tunneling process, we apply the classical WKB method to estimate the Hawking radiation temperature.

### 2.1 Phonon emission from the spherically symmetric sonic black hole

For spherically symmetric convergent stationary flow, the effective metric in the line element form is [25]

$$ds^2 = g_{\mu\nu} dx^\mu dx^\nu = \frac{\rho_0}{c_s}\left[-\left(c_s^2 - v_r^2\right)dt^2 + 2v_r dr\, dt + dr^2 + r^2\left(d\theta^2 + \sin\theta^2 d\varphi^2\right)\right], \quad (1)$$

where $\rho_0$ denotes the density of the background flow and the positive sign in the right second term relates to the convergent assumption. One can find that this is a natural Painlevé-like coordinate system in which the necessary conditions for semi-classical approximation are satisfied [8]. In the WKB zero order approximation, for stationary spactime, the quantized sound wave function becomes

$$\Psi(\vec{x}, t) \sim e^{i/\hbar S(\vec{x})} = e^{i/\hbar[-Et + S_0(\vec{x})]} = e^{i/\hbar[-Et + \int \vec{p}\, d\vec{x}]}, \quad (2)$$

where $E$ is finite phonon energy, $S(\vec{x})$ is the quasi-particle action and $\vec{p}$ is the spatial part of covariant relativistic momentum which satisfies

$$\nabla S_0(\vec{x}) = \vec{p} \quad (3)$$

and

$$g^{00} E^2 - 2g^{0i} E p_i + g^{ij} p_i p_j + \beta m^2 c_s^2 = 0 \quad (4)$$

($\beta$ is the conformal factor, specifically i.e. $\beta = \rho_0/c_s$, which influences only the left 4-th term, and $m$ represents mass, i.e. massive phonon [26], but zero in current case, i.e. massless phonon; this zero order approximation equation is independent of time and different from that of the true particle). In 0-th approximation $S(\vec{x},t) = -Et + S_0(\vec{x})$, the negative sign before $E$ plays a role. Because in quantum mechanics, the Hamiltonian $H$ represents the energy $E$, and the solution of the 0-th approximation is relative to the Hamilton action $S$. If the quantum system is apparently time independent, there will be $\frac{\partial S}{\partial t} = -H = -E$. So in Hamilton action $S$, here the energy must have negative sign for consistence.

The radial part of the momentum is responsible for phonon tunneling process, whilst the azimuthal parts are independent to it, therefore being ignored. So the momentum can be written as

$$\vec{p} = (p_r, 0, 0), \tag{5}$$

then from Eq.(1), Eq.(4) and Eq.(5), noting that $m = 0$, we get

$$p_r = E\frac{\upsilon_r \pm c_s}{c_s^2 - \upsilon_r^2} = E\frac{1}{\pm c_s - \upsilon_r}. \tag{6}$$

The classical tunneling probability across the horizon for phonon is

$$\Gamma \sim |\Psi|^2 = e^{-2/\hbar \operatorname{Im} \int p_r \, dr}. \tag{7}$$

We choose the radial momentum with the sign $+$ in Eq.(6) and obtain

$$\Gamma \sim e^{-2/\hbar \operatorname{Im} \int p_r \, dr} = e^{-2/\hbar \operatorname{Im} \int_{r_i}^{r_f} E \frac{1}{c_s - \upsilon_r} dr} = e^{-2\pi E/\hbar\alpha}, \tag{8}$$

where $r_i$ is just inside the horizon and $r_f$ is just outside, and we have assumed $c_s - \upsilon_r$ has zero point of order 1 at the horizon $c_s = \upsilon_r$ which is usually physical. Compare Eq.(8) with Boltzmann factor $\Gamma \propto e^{-E/k_B T}$, and one can read out the Hawking temperature

$$T_H = \frac{\hbar\alpha}{2\pi k_B}. \tag{9}$$

The result is consistent with that in other methods [25, 19] and independent of the conformal factor $\beta$.

## 2.2 In stationary cylindrically symmetric sonic black hole case

For identifying the effectiveness of the classical WKB method in the phonon tunneling process and providing cases for subsequent discussion, we calculate the Hawking radiation from stationary rotational sonic black hole. For stationary cylindrically symmetric convergent flow, the line element turns [27]

$$ds^2 = \frac{\rho_0}{c_s}\left[-\left(c_s^2 - v_r^2 - v_\varphi^2\right)dt^2 + 2v_r\,dr\,dt + dr^2 - 2v_\varphi r\,d\varphi\,dt + r^2\,d\varphi^2 + dz^2\right], \quad (10)$$

where $c_s, \rho_0, v_r, v_\varphi$ are only change with $r$ and the sign + is relative to the convergent feature.

Assume that the flow velocity is transonic, then the sonic locates at $r = a$ where $v_r = c_s$. This coordinate system is also Painlevé-like. We write the spatial part of covariant relativistic momentum as

$$\vec{p} = (p_r, p_\varphi, 0) = (p_r, j\hbar, 0), \quad (11)$$

where $p_\varphi$ is the general momentum relative to $\varphi$ coordinate, i.e. angular momentum $j\hbar$ of tunneling phonon ($j$ is an integer). Substitute Eq.(11) into Eq.(4), and for massless phonon, one can get

$$p_r = \frac{v_r\left(E - \frac{v_\varphi p_\varphi}{r}\right) \pm \sqrt{c_s^2\left(E - \frac{v_\varphi p_\varphi}{r}\right)^2 - \left(c_s^2 - v_r^2\right)\frac{c_s^2 p_\varphi^2}{r^2}}}{c_s^2 - v_r^2}, \quad (12)$$

then the tunneling probability

$$\Gamma \sim e^{-2/\hbar \operatorname{Im}\int p_r\,dr + p_\varphi\,d\varphi} = e^{-2/\hbar \operatorname{Im}\int p_r\,dr + j\hbar\,d\varphi}. \quad (13)$$

Choose the momentum with sign + in Eq.(12) which notates "outgoing", so the integral becomes

$$\operatorname{Im}\int p_r\,dr = \operatorname{Im}\int_{r_i}^{r_f} p_r\,dr = \frac{\pi(E - j\hbar v_\varphi/a)}{\alpha} = \frac{\pi(E - \omega_0 j\hbar)}{\alpha}, \quad (14)$$

where $\omega_0$ is the proper angular velocity of sonic black hole, while

$$\operatorname{Im}\int p_\varphi\,d\varphi = \operatorname{Im}\int j\hbar\,d\varphi = \operatorname{Im}\int_{\varphi_i}^{\varphi_f} j\hbar\,d\varphi = 0. \quad (15)$$

Inserting Eq.(14) and Eq.(15) into Eq.(13), the result is

$$\Gamma \sim e^{-2\pi(E - \omega_0 j\hbar)/\hbar\alpha}. \quad (16)$$

One can find the phonon emission temperature is similar with Eq.(9). For the "draining bathtub" sonic spacetime [25], $v_r = A/r, v_\varphi = B/r$ ($A, B$ are constant), the horizon is at $r = a = A/c_s$, $\omega_0 = \left(v_\varphi|_{r=a}\right)/a = Bc_s^2/A^2$, $\alpha = c_s^2/A$, so we get the sonic Hawking temperature

$$T_H = \frac{\hbar c_s^2}{2\pi k_B A} \quad (17)$$

and the tunneling probability

$$\Gamma \sim e^{-2\pi A\left(E - \hbar Bc_s^2/A^2\right)/\hbar c_s^2} = e^{2\pi\left(B/A - EA/\hbar c_s^2\right)}. \quad (18)$$

In the analysis above, we show that the classical WKB method used in Painlevé-like coordinates can give the expected results in the phonon tunneling process. Note that we use covariant relativistic momentum to calculate the imaginary part of the quasi-particle action above. Instead, without regard to the backreaction, we can also use the Hamilton equation

$$\frac{\partial H}{\partial p_r} = \dot{r}, \frac{\partial H}{\partial p_\varphi} = \dot{\varphi}, \frac{\partial H}{\partial r} = \dot{p}_r = \frac{\partial H}{\partial \varphi} = \dot{p}_\varphi = \frac{\partial H}{\partial t} = 0 \qquad (19)$$

(the dot denotes derivative in respect to time) to simplifies the integrals

$$\mathrm{Im} \int p_r \, \mathrm{d}r = \mathrm{Im} \int \mathrm{d}r \int \mathrm{d}p_r = \mathrm{Im} \int \mathrm{d}r \int \frac{\mathrm{d}H}{\dot{r}} = \mathrm{Im} \int \mathrm{d}r \int \frac{\mathrm{d}E}{\dot{r}} \qquad (20)$$

and

$$\mathrm{Im} \int p_r \, \mathrm{d}r + p_\varphi \, \mathrm{d}\varphi = \mathrm{Im} \int \mathrm{d}r \int \mathrm{d}p_r + \int p_\varphi \, \mathrm{d}\varphi$$
$$= \mathrm{Im} \int \mathrm{d}r \left( \int \frac{\mathrm{d}H}{\dot{r}} - \int \frac{\dot{\varphi}}{\dot{r}} \mathrm{d}p_\varphi \right) + \int p_\varphi \, \mathrm{d}\varphi = \mathrm{Im} \int \mathrm{d}r \int \frac{\mathrm{d}E}{\dot{r}} - p_\varphi \int \frac{\dot{\varphi}}{\dot{r}} \mathrm{d}r. \qquad (21)$$

Let $\mathrm{d}s^2 = \mathrm{d}\theta^2 = \mathrm{d}\varphi^2 = 0$ in Eq.(1), we have

$$\dot{r} = -\upsilon_r \pm c_s. \qquad (22)$$

For the outgoing phonon, one can get

$$\mathrm{Im} \int p_r \, \mathrm{d}r = \mathrm{Im} \int_{r_i}^{r_f} \mathrm{d}r \int_0^E \frac{\mathrm{d}E}{-\upsilon_r + c_s} = \mathrm{Im} \int_{r_i}^{r_f} \frac{E}{-\upsilon_r + c_s} \mathrm{d}r = \frac{\pi E}{\alpha}. \qquad (23)$$

Let $\mathrm{d}s^2 = \mathrm{d}z^2 = 0$ in Eq.(10) to yield

$$\dot{r} = -\upsilon_r \pm \sqrt{c_s^2 - (\upsilon_\varphi/r - \dot{\varphi})^2 / r^2}, \qquad (24)$$

take the one with sign + and note that $\frac{\upsilon_\varphi}{r} = \omega_0$ at the horizon, one can get

$$\mathrm{Im} \int p_r \, \mathrm{d}r + p_\varphi \, \mathrm{d}\varphi$$
$$= \mathrm{Im} \int_{r_i}^{r_f} \mathrm{d}r \int_0^E \frac{\mathrm{d}E}{-\upsilon_r + \sqrt{c_s^2 - (\upsilon_\varphi/r - \dot{\varphi})^2 / r^2}} - p_\varphi \int_{r_i}^{r_f} \frac{\dot{\varphi}}{-\upsilon_r + \sqrt{c_s^2 - (\upsilon_\varphi/r - \dot{\varphi})^2 / r^2}} \mathrm{d}r. \qquad (25)$$
$$= \mathrm{Im} \int_{r_i}^{r_f} \frac{E - p_\varphi \dot{\varphi}}{-\upsilon_r + \sqrt{c_s^2 - (\upsilon_\varphi/r - \dot{\varphi})^2 / r^2}} \mathrm{d}r = \frac{\pi(E - \omega_0 j\hbar)}{\alpha}$$

One can see that the result is consistent and the derivation is further simplified.

# 3. Modified phonon tunneling calculation

Although the classical WKB approximation works well in Painlevé-like coordinates, the fact that it is variable in canonical transform results in the question to it. Chowdhury proposed the closed path integral [15]

$$\Gamma \sim e^{-1/\hbar \, \mathrm{Im} \oint p_r \, \mathrm{d}r}, \quad (26)$$

which keeps invariant in canonical transform. But, in many cases, the closed path integral gives twice larger Hawking temperature. The WKB tunneling probability is often considered only from the spatial integral before, and Akhmedov, Akhmedova et al. pointed out the temporal part also contributes to it because the features of the space and time coordinates exchange each other when crossing the horizon. Taking into account of the temporal contribution, the tunneling probability becomes [16, 17]

$$\Gamma \sim e^{1/\hbar \, \mathrm{Im}\left(E\Delta t^{\mathrm{out}} + E\Delta t^{\mathrm{in}} - \oint p_r \, \mathrm{d}r\right)}. \quad (27)$$

where $\Delta t^{\mathrm{out}}$, $\Delta t^{\mathrm{in}}$ are the differences of time across the horizon. This is the WKB loop (closed path) integral. Now we apply it to sonic black hole case.

About the temporal contribution, the trick is to find the change from time coordinate on one side of the horizon to space coordinate on the other side, usually by the coordinate transform [17]. We take that as a proper feature of the system, one can find the temporal contribution by the velocity integral of null geodesics without the coordinate transform. Now that the temporal contribution comes from the exchange of space-time across the horizon, we can obtain it by the space-time relationship (i.e. describing the time $t$ coordinate with the space coordinate $r$) for massless particle if possible. In fact, by integrating the velocity along the null geodesics, one can get the space-time relationship in principal. In spherically symmetrical stationary flow case, the velocity of null geodesics is shown in Eq.(22), which can be rearranged as

$$\frac{\mathrm{d}r}{-\upsilon_r \pm c_s} = \mathrm{d}t. \quad (28)$$

Integrate Eq.(28) over the horizon, one can obtain

$$\Delta t^{\mathrm{out}} = \int_{t^-}^{t^+} \mathrm{d}t = \int_{R^-}^{R^+} \frac{\mathrm{d}r}{-\upsilon_r + c_s} \approx \int_{R^-}^{R^+} \frac{\mathrm{d}r}{\alpha(r-R)} = \frac{1}{\alpha}\ln(r-R)\Big|_{R^-}^{R^+} = -\frac{i\pi}{\alpha} + \mathrm{realpart} \quad (29)$$

and

$$\Delta t^{\mathrm{in}} = \int_{t^+}^{t^-} \mathrm{d}t = \int_{R^+}^{R^-} \frac{\mathrm{d}\upsilon_r}{-\upsilon_r - c_s} = i0 + \mathrm{realpart}, \quad (30)$$

where $t^-, t^+$ are time and $R^-, R^+$ are radial position just inside and outside the horizon in the tunneling process. The radial momentum of phonon is shown in Eq.(6), so

$$\oint p_r \, \mathrm{d}r = \int_{R^-}^{R^+} p^{\mathrm{out}} \, \mathrm{d}r - \int_{R^+}^{R^-} p^{\mathrm{in}} \, \mathrm{d}r = \int_{R^-}^{R^+} \frac{E \, \mathrm{d}r}{-\upsilon_r + c_s} - \int_{R^+}^{R^-} \frac{E \, \mathrm{d}r}{-\upsilon_r - c_s} = \frac{i\pi E}{\alpha} + \mathrm{realpart}. \quad (31)$$

and the tunneling probability is

$$\Gamma \sim e^{1/\hbar \, \mathrm{Im}\left(E\Delta t^{\mathrm{out}}+E\Delta t^{\mathrm{in}}-\oint p_r \, \mathrm{d}r\right)} = e^{1/\hbar(-\pi E/\alpha + 0 - \pi E/\alpha - 0)} = e^{-2\pi E/\hbar\alpha} . \tag{32}$$

In cylindrically symmetrical stationary convergent flow case, the momentum is relative to Eq.(12), and we have

$$\oint \left(p_r \, \mathrm{d}r + p_\varphi \, \mathrm{d}\varphi\right) = \oint p_r \, \mathrm{d}r = \int_{R^-}^{R^+} p_r^{\mathrm{out}} \, \mathrm{d}r - \int_{R^+}^{R^-} p_r^{\mathrm{in}} \, \mathrm{d}r = \frac{i\pi E}{\alpha} + \mathrm{real\,part} . \tag{33}$$

Note that $\int p_\varphi \, \mathrm{d}\varphi = j\hbar\varphi$ and quasi-particle produced has angular velocity $\omega_0$ at the horizon, i.e. $\varphi = \omega_0 t$, the Hamilton action becomes

$$S = -Et + \int p_r \, \mathrm{d}r + j\hbar\varphi = -(E - j\hbar\omega_0)t + \int p_r \, \mathrm{d}r . \tag{34}$$

To find the temporal contribution, we make use of Eq.(24) to get

$$\Delta t^{\mathrm{out}} = \int_{t^-}^{t^+} \mathrm{d}t = \int_{a^-}^{a^+} \frac{\mathrm{d}r}{-v_r + \sqrt{c_s^2 - (v_\varphi/r - \dot{\varphi})^2/r^2}} \approx \int_{a^-}^{a^+} \frac{\mathrm{d}r}{\alpha(r-a)} = -\frac{i\pi}{\alpha} + \mathrm{real\,part} \tag{35}$$

and

$$\Delta t^{\mathrm{in}} = \int_{t+}^{t^-} \mathrm{d}t = \int_{a^-}^{a^+} \frac{\mathrm{d}r}{-v_r - \sqrt{c_s^2 - (v_\varphi/r - \dot{\varphi})^2/r^2}} = i0 + \mathrm{real\,part} . \tag{36}$$

Then we get the tunneling probability

$$\Gamma \sim e^{1/\hbar \, \mathrm{Im}\left((E-j\hbar\omega_0)(\Delta t^{\mathrm{out}}+\Delta t^{\mathrm{in}})-\oint p_r \, \mathrm{d}r\right)} = e^{-\pi(E-j\hbar\omega_0)/\hbar\alpha + 0 - \pi(E-j\hbar\omega_0)/\hbar\alpha - 0} = e^{-2\pi(E-j\hbar\omega_0)/\hbar\alpha} . \tag{37}$$

We again obtain the expected results (see Eq.(32) and Eq.(37)) in the modified WKB approximation method, from which the way to find the temporal contribution by null geodesics velocity integral is confirmed to be effective.

## 4. Same results in Painlevé-like coordinates

From the calculation above and other analysis in the classical WKB formula, one can find the method is effective and there is no exceptional failure within the range of approximation. In fact, by making formal (symbolized) calculation, in (sonic) black hole case, one can find that the results are the same with each other in Painlevé-like coordinates.

From Eq.(7) and Eq.(27) or Eq.(13) and Eq.(37), it is easy to find that the differences from the two formulae are the temporal contribution and the spatial contribution. If they are proved to be equal, then the results from the two formulae become the same.

We now consider the stationary 1-D case, the line element in Painlevé-like coordinates is

$$\mathrm{d}s^2 = g_{00} \, \mathrm{d}t^2 + 2g_{01} \, \mathrm{d}r \, \mathrm{d}t + g_{11} \, \mathrm{d}r^2 . \tag{38}$$

Using the 0-th approximation equation (similar to Eq.(4)), the radial momentum for massless particle is

$$p_r^\pm = \frac{E\left(-g_{01}/g \mp \sqrt{g_{01}^2/g^2 - g_{00}g_{11}/g^2}\right)}{g_{00}/g} = \frac{E\left(-g_{01} \mp \sqrt{g_{01}^2 - g_{00}g_{11}}\right)}{g_{00}}, \quad (39)$$

where $g = \det(g_{\mu\nu}) < 0$ in (sonic) black hole case. By the condition of $ds^2 = 0$, we get the velocity along null geodesics

$$v_r^\pm = \frac{-g_{01} \pm \sqrt{g_{01}^2 - g_{00}g_{11}}}{g_{11}} = \frac{g_{00}}{-g_{01} \mp \sqrt{g_{01}^2 - g_{00}g_{11}}} = \frac{E}{p_r^\pm}. \quad (40)$$

So we have

$$Et = E\int dt = E\int \frac{dr}{v_r} = E\int \frac{dr}{E/p_r} = \int p_r \, dr, \quad (41)$$

and

$$\operatorname{Im} E\Delta t^{\text{out,in}} = -\operatorname{Im} \int p_r^{\text{out,in}} \, dr \quad (42)$$

(similar to Eq.(29), derivative expansion and approximate integral). From Eq.(39), one can find the imaginary part of the momentum integral involves the zero point (generally of order 1) of $g_{00}$ at the horizon and only one part of the $\pm$ gives a nonzero imaginary part(in (sonic) white hole case, the barrier is loaded on the ingoing mode with sign $-$ and the part with sign $+$ contributes zero to the imaginary part). Thus, one can get

$$\begin{aligned}\operatorname{Im}\left(E\Delta t^{\text{out}} + E\Delta t^{\text{in}} - \oint p_r \, dr\right) &= \operatorname{Im}\left(-\int p_r^{\text{out}} \, dr + \int p_r^{\text{in}} \, dr - \oint p_r \, dr\right) \\ &= -2\operatorname{Im}\oint p_r \, dr = \mp 2\operatorname{Im}\int_{\text{horizon}^-}^{\text{horizon}^+} p_r^{\text{out,in}} \, dr\end{aligned}, \quad (43)$$

where sign $\mp$ is relative to "out" and "in". Eq.(43) manifests the same results will be got whatever formula one chooses in Painlevé-like coordinates.

In cylindrically symmetric stationary 2-D case, the square infinitesimal distance is

$$ds^2 = g_{00} \, dt^2 + 2g_{01} \, dr\,dt + g_{11} \, dr^2 + 2g_{02} \, d\varphi\,dt + g_{22} \, d\varphi^2, \quad (44)$$

and we assume the existence of horizon. The equation for the 0-th approximation solution is similar to Eq.(4), so we obtain the radial momentum

$$p_r^\pm = \frac{-E(g_{01}g_{22} - g_{02}g_{12}) + (g_{00}g_{12} - g_{01}g_{02})p_\varphi \mp \sqrt{-gg_{22}\left(E + p_\varphi g_{02}/g_{22}\right)^2 - g\left(g_{00} - g_{02}^2/g_{22}\right)p_\varphi^2}}{g_{00}g_{22} - g_{02}^2}. \quad (45)$$

The term $g_{00} - g_{02}^2/g_{22} = 0$ at the horizon, and one can confirm it immediately by the transform $d\varphi = \frac{g_{02}}{g_{22}} dt + d\varphi$ to eliminate the parameter $\varphi$. In Painlevé-like coordinates, the metric components are regular, so the imaginary part of momentum integral comes from the zero point of the denominator. Obviously, either the ingoing momentum or the outgoing momentum contributes to the imaginary part of the momentum integral, and the other one contributes zero imaginary part because of the limit of type $0/0$. At the horizon, we have the identity

$$\sqrt{-gg_{22}} = \sqrt{(g_{01}g_{22} - g_{02}g_{12})^2}, \qquad (46)$$

and Eq.(45) becomes

$$p_r^{\pm} = \frac{\sqrt{-gg_{22}}(E + p_{\varphi}g_{02}/g_{22})(1 \pm 1)}{-(g_{00}g_{22} - g_{02}^2)}. \qquad (47)$$

The particle velocity along null geodesics is

$$v_r^{\pm} = \frac{-(g_{01} + g_{12}\dot{\varphi}) \pm \sqrt{(g_{01} + g_{12}\dot{\varphi})^2 - g_{11}(g_{00} + 2g_{02}\dot{\varphi} + g_{22}\dot{\varphi}^2)}}{g_{11}}, \qquad (48)$$

and substitute $\dot{\varphi}$ with $\dot{\varphi} = \omega_0 = -g_{02}/g_{22}$ at the horizon into Eq.(48), with Eq.(46), we get

$$\frac{1}{v_r^{\pm}} = \frac{-(g_{01}g_{22} - g_{02}g_{12}) \mp \sqrt{-gg_{22}}}{g_{00}g_{22} - g_{02}^2} = \frac{\sqrt{-gg_{22}}(1 \pm 1)}{-(g_{00}g_{22} - g_{02}^2)}. \qquad (49)$$

Note that we have considered that quasi-particle produced at the horizon has angular velocity $\omega_0$ and angular momentum $p_{\varphi_0}$, from Eq.(49) and Eq.(47), one can obtain the temporal difference across the horizon

$$(E - p_{\varphi_0}\omega_0)\Delta t = (E - p_{\varphi_0}\omega_0)\int dt = (E - p_{\varphi_0}\omega_0)\int \frac{dr}{v_r} = \int p_r \, dr. \qquad (50)$$

With regard to Eq.(42), we get

$$\text{Im}\left((E - j\hbar\omega_0)(\Delta t^{\text{out}} + \Delta t^{\text{in}}) - \oint p_r \, dr\right) = -2\,\text{Im}\oint p_r \, dr = \mp 2\,\text{Im}\int p_r^{\text{out,in}} \, dr. \qquad (51)$$

We can see that, in Painlevé-like coordinates, the two momentum integral formulae give the same results in both 1-D and cylindrically symmetric 2-D stationary spacetime cases. This shows that the classical momentum integral from WKB approximation is reliable to put into use on the tunneling process in Painlevé-like coordinates.

## 5. Conclusion

We use both the classical WKB approximation and the modified WKB approximation (WKB loop integral) to calculate the tunneling probability of quasi-particle (phonon) crossing the stationary sonic horizon. The both results obtained are accord with the usual ones [25], which in return show that our work is consistent. We prove that the results from the two tunneling probability formulae are completely equivalent in Painlevé-like coordinate system in some cases. The classical WKB approximation is more intuitive and easy to construct the tunneling process in a uniformed picture. It is obvious that the modified tunneling probability formula is further stricter in theory because of the canonical invariant, though the concerned temporal contribution is not well understood and overlook in convention. It seems to be no limits on the regular requirement of the metric at the horizon, and whether it is applicable or not in singular metric case is worth to further study.

## Acknowledgement

This research is supported by the Natural Science Foundation of China (10847166).